\documentstyle[11pt,aaspp4]{article}
\begin{document}

\title{The IntraCluster Medium: An Invariant Stellar IMF}

\author{Rosemary F.~G.~Wyse\altaffilmark{1}}
\affil{Dept. of Physics and Astronomy, The Johns Hopkins University, 
Bloomberg Center, 
Baltimore, MD 21218 \\}

\altaffiltext{1}{wyse@tardis.pha.jhu.edu}

\begin{abstract} 
Evidence supporting the hypothesis of an invariant stellar Initial Mass
Function is strong and varied.  The intra-cluster medium in rich clusters of
galaxies is one of the few contrary locations where recent interpretations of
the chemical abundances have favoured an IMF that is biased towards massive
stars, compared to the `normal' IMF.  This interpretation hinges upon the
neglect of Type Ia supernovae to the ICM enrichment, and a particular choice of
the nucleosynthesis yields  of Type II supernovae. We demonstrate here that
when one adopts  yields determined empirically from observations of Galactic
stars, rather than the uncertain model yields, a self-consistent picture may be
obtained with an invariant stellar IMF, and about half of the iron in the ICM
being produced by Type~Ia supernovae.
\end{abstract}

\section{Introduction}

In cases when one can  count stars reliably, the derived stellar Initial Mass
Function (IMF) is remarkably invariant -- e.g. low mass stars in globular and
open clusters in the Milky Way Galaxy, spanning two orders of magnitude in
metallicity (von Hippel {\it et al.} 1996); low mass stars in the field halo
(Chabrier and Mera 1997) and disk  (Kroupa, Tout and Gilmore 1993); young
star-forming regions in the Galactic disk (Lada 1997), and even for the
intermediate -- massive stars in the cluster R136a of our local `starburst'
region 30 Doradus (Hunter {\it et al.} 1995).  Further, even in more exotic
locations,  such as the starburst galaxy M82 where indirect inferences of the
IMF must be made, there is currently no significant evidence for an anomalous
IMF (Leitherer 1997).  For massive stars, this universal IMF is reasonably
modelled by a power law with slope similar to that of Salpeter (1955).

The  elemental abundances -- both relative and absolute values --
of the gas in the Intra Cluster Medium (ICM) have been interpreted  as showing 
the gas to have been
enriched by stars with a biased IMF, being top-heavy with respect to
the `normal', local IMF (e.g. Loewenstein and Mushotzky 1996). Their 
derived elemental ratios implied that only Type II supernovae
enriched the ICM, and the actual elemental 
abundances were too high for this to
be achieved with a normal IMF.  However, uncertainties in theoretical yields
from supernovae complicate the interpretation of these data by comparison with
models (e.g. Gibson, Loewenstein and Mushotzky 1997).  An empirical comparison
with elemental abundances of objects with reasonably well-understood chemical
evolution offers an alternative approach.  
The main result of this letter is given in Figure~1, which shows elemental 
data for Galactic disk stars, together with the ICM data. The ICM points fall
within those of the stars, leading to the conclusion that the simple
picture of an invariant IMF, that of the Milky Way Galaxy, 
is equally consistent with the data. There is no
compelling need to invoke special conditions.

\section{Chemical Enrichment Constraints on the IMF}

Figure~1 here shows element ratios with respect to iron, versus 
the iron abundance, for local Galactic disk 
stars and for the ICM.  The combination of 
element ratios and overall abundance is a powerful constraint on the stellar 
IMF. 

\subsection{Element Ratios}

Indirect constraints on the stellar IMF may be obtained by study of
the elemental abundances of stars and of gas, exploiting the fact that
different elements are synthesized by stars of different mass.  In
particular, oxygen and the alpha-elements (created by the addition of
successive $^4$He nuclei) are formed overwhelmingly in core-collapse
Type II supernovae, the endpoints of stars more massive than ${M_{II}
\sim 8 M_\odot}$, while iron has a significant production in {\it
both\/} Type II and Type Ia supernovae. While the understanding of
Type Ia supernovae is by no means complete, models in general invoke
explosive nucleosynthesis in a white dwarf driven over the
Chandrasekhar mass limit by accretion from a binary companion
(e.g. Iben and Tutukov 1987).  The progenitor stars are of
intermediate to low mass, and the final explosion may occur long after
the formation of the white dwarf(s), determined by orbital parameters
of the binary.  Thus Type~II supernovae explode shortly after the
progenitor massive stars are born, releasing lots of oxygen, while the
onset of the ejection of significant amounts of enriched material from
Type Ia supernovae is delayed, probably for the order of 1~Gyr, and continues
long after any cessation of star formation (e.g. Wheeler, Sneden and
Truran 1989; Gilmore, Wyse and Kuijken 1989).

 Stars formed in the early stages of chemical evolution -- such as
expected for the metal-poor stars in the Milky Way -- will form from
gas enriched only by Type II supernovae, and are expected to be
relatively rich in oxygen and the alpha-elements, but poor in
iron. The short timescales on which all Type II explode lead (with efficient 
mixing) to the
expectation that one should observe a (massive-star) 
IMF-averaged [O/Fe] value.  The
exact value of the ratio of oxygen to iron depends on the massive star
IMF, since the oxygen yield of a Type II supernova is a steeply
increasing function of main sequence mass, while the iron yield is
approximately independent of progenitor mass (e.g.~Thielemann  {\it
et al.} 1993).  However, as noted by Gibson,
Loewenstein and Mushotzky (1997), there is currently a
factor of 2--3 range in the available model predictions of the
IMF-averaged elemental yields of Type~II supernovae, even with IMF
fixed.  Hence a determination of the actual value of a parameter of 
the IMF, such as the slope,  is
not reliable from element ratio data alone, but a very robust
constraint on any possible {\it variation\/} in the 
value of this parameter of 
the massive star IMF may be derived
from any variation of the observed elemental abundances in objects
enriched by Type II supernovae alone.  Stars and gas pre-enriched by
an IMF biased towards higher mass stars will have a higher value of
oxygen-to-iron (e.g. Wyse and Gilmore 1992).

The rate of decline in O/Fe (for example) that occurs once Type Ia
supernovae commence depends on the relative rates of Type II and Type
Ia supernovae, and thus on the present star formation rate, and star
formation history. Obviously the steepest decline occurs when star
formation ceases totally, and only Type Ia supernovae are contributing
to chemical enrichment, providing negligible oxygen but significant
iron.  A slowly decreasing star formation rate with time (as inferred
for the local Galactic disk), and normal IMF, provides a somewhat more
gentle decline.  A plateau in oxygen-to-iron, reflecting enrichment by 
Type~II supernovae alone, followed by a decline in oxygen-to-iron ratio, is
a robust prediction for non-pathological star-formation histories (see
Gilmore and Wyse 1991 for variations).

\subsection{Overall Enrichment}

The overall level of chemical enrichment achieved in a system of given
gas fraction depends on the stellar IMF, and thus in principle by itself 
provides a constraint.  However, it is often difficult to estimate how
much of the gas is actually taking part in star formation, and the
(probable) occurence of gas flows further complicates the
interpretations (e.g.~Edmunds 1990).  In the present context, for any
fixed IMF and specific set of model yields, one can predict an
expected abundance of various elements if only Type~II supernovae
enrich the medium.  A {\it higher\/} abundance, for fixed yields,
would mean either biasing of the IMF towards more massive stars (e.g.
Loewenstein and Mushotzky 1996), or a significant contribution from
Type Ia supernovae (obviously for those elements produced by these).
The element ratio constraint then enters, since as noted above, such a biased 
IMF would increase the alpha-element to iron ratio, while allowing Type Ia 
supernovae would decrease the alpha-element to iron ratio. 

\section{Observational Input}

\subsection{The IntraCluster Medium}

Estimates of the iron abundance in the hot intergalactic medium in
clusters of galaxies showed it to be rather enriched, with a typical
metallicity of one-third solar (Arnaud {\it et al.} 1992).  Data from
the ASCA satellite has allowed additional elemental abundances in the
ICM in four rich clusters of galaxies to be estimated (Mushotzky {\it et
al.} 1996).   Understanding  these results is
facilitated by comparing them to samples of stars within which one can
establish trends.\footnote{Ideally one would do this for stars in the
dominant galaxies in clusters, the ones that presumably enriched the
IGM, if one had the observational data.  However, the comparison is
only possible with stars in our Galaxy, for which there is a good
dataset and a reasonable understanding of the chemical evolution that
has determined the variation of the elemental abundances.}  Direct
comparison of these ICM elemental abundances with the data for
Galactic stars is possible once the ICM results are put on the same
elemental abundance scale as the stars. For the stellar datasets
discussed here this is based on 
the meteoretic iron abundance of Anders and Grevesse
(1989), which agrees with the solar photospheric iron abundance of
e.g. Holweger {\it et al.} (1991) and Biemont {\it et al.} (1991). 
As recently discussed by Ishimaru and Arimoto (1997) and by
Gibson, Loewenstein and Mushotzky (1997), this calibration results in
mean values of the ICM elemental abundance ratios of [O/Fe] $\sim
+0.0$ (i.e. solar ratio); [Si/Fe]$ \sim +0.13$; [Mg/Fe] $\sim -0.1$.
The mean iron abundance of this gas is [Fe/H]$\sim -0.3$
(re-calibrated from Mushotzky {\it et al.} 1996 who used a higher
solar iron abundance, by $\sim 0.2$~dex). The individual re-calibrated 
ratios of oxygen and of 
silicon relative to 
iron, and iron abundance, 
for the ICM in the four clusters analysed by Loewenstein and Mushotzky (1996) 
are displayed in
Figure~1.  These elements are representative of those studied in stars and in 
X-ray gas. 

\subsection{Galactic Stars}

Abundance analyses based on high-resolution spectroscopy have shown
that essentially all metal-poor halo stars (observed locally at the
solar neighbourhood) have the same value of elemental ratios, with
[$\alpha$/Fe]$ \sim +0.3$ (oxygen is the element with the
largest uncertainties).  Indeed, the lack of scatter in alpha-to-iron
ratio seen in samples of halo stars, down to metallicities [Fe/H]$
\sim -3$ dex, places severe constraints on any possible variation in
the slope of the massive-star IMF (although as noted above, one must adopt 
theoretical yields to be able to say what is 
the actual value of the slope).  For example, adopting Type II
yields from Thielemann {\it et al.} (1992), Nissen {\it et al.} (1994)
determined that the allowable range of slopes of a power-law IMF was
at most $ 0.8 < x < 1.8$ (one-sigma range), where the familiar
Salpeter slope is $x=1.35$.  

As shown in Figure~1, the same value of the
`Type II plateau' is seen in local metal-poor (thick) disk stars, decreasing
(due to the addition of Type Ia iron) with increasing metallicity above
[Fe/H]$ \sim -0.6$, reaching approximately solar values of the element
ratios at approximately solar iron (Edvardsson {\it et al.}  1993). 

Note that the Edvardsson {\it et al.} data show a large, real,  
scatter in the age-metallicity relationship, and in particular that the Sun is
rather more metal-rich than the average for its age.  Indeed, recent work has
established that the B stars in the Orion Association have the same metallicity
as the Orion Nebula, and thus are more metal-poor than is the Sun, [O/H]$ \sim
-0.3$~dex (Gies and Lambert 1992; Cunha and Lambert 1994).

Timmes, Woosley and Weaver (1995) undertook a comprehensive investigation of 
the evolution of elemental abundances in the Galaxy, as probed by local stars
(including high velocity halo stars) over the entire metallicity range from
$-3$~dex to solar. They included in their discussion essentially every element
with both predictions and observations; their  
model assumed a quadratic Schmidt law of star formation, a fixed (Salpeter) 
IMF slope, the Type
II yields of Woosley and Weaver (1995), and standard Type Ia yields from
Thielemann, Nomoto and Yokoi (1986).  Timmes {\it et al.} 
could understand the behavior 
of the element
ratios with respect to iron, as a function of iron abundance, 
for a remarkably wide range of elements.
The element ratio trends,  
combined with
the overall age--iron abundance data, are consistent with one-third to
two-thirds of the iron in the solar neighbourhood now having been contributed
by Type Ia supernovae.

Thus the observations of Galactic stars require that the massive star IMF be
invariant over $10^{2.5}$ range in iron abundance, from the lack of scatter 
about the Type II plateau in element ratios,  and
are consistent with no variation of the IMF or of the yields 
for any star formation region
anywhere, any time, in the Galaxy.

\section{An Invariant IMF}

Figure 1 here shows the ICM points lying within those of Galactic
stars.\footnote{Note that one should not expect exact agreement
between the Galactic stars and the ICM, since element ratios at a
given iron abundance depend not only on IMF but on star formation
history, and this may well be different for the gas in the ICM now and
in the Milky Way disk. However, the excellent agreement here may 
well support theories, such as hierarchical-clustering Cold Dark Matter 
dominated models, in which a very significant fraction of the metals in the 
ICM are produced in disk galaxies, precursors to giant ellipticals 
(e.g. Kauffmann and Charlot 1997).}
 Thus from the discussion above, one
might expect there to have been a significant contribution of Type Ia
supernovae also for the ICM. Indeed, the results of Gibson,
Loewenstein and Mushotzky (1997) imply that adopting the Woosley and
Weaver yields, the elemental abundances of oxygen and silicon relative
to iron in the ICM do imply that around 0.4 of the ICM iron originated
in Type~Ia supernovae (their Figure 1).  This was the conclusion of
Ishimaru and Arimoto (1997), on the basis of the ICM data alone, but
contested by Gibson, Loewenstein and Mushotzky (1997). These latter
authors emphasise that it is possible to obtain Type II yields that have
approximately solar element ratios by themselves, without any addition of 
iron from Type~Ia supernovae.  To illustrate, Gibson {\it et al.} calculated
approximately solar oxygen-to-iron element ratios from the Type~II
supernova models of Maeder (1992) -- these invoke
efficient pre-supernova mass-loss --  and those of Arnett (1996) 
-- these invoke very efficient convection.  Thus Gibson {\it et al.}
propose that the ICM data remain consistent with {\it pure\/} Type~II
enrichment, and thus the overall level of enrichment of the ICM would then
still require a top-heavy IMF. 

However, there is the further constraint discussed here, in that those 
Type~II 
yields do {\it not\/} provide an explanation for the Galactic star
data discussed above, since they fail to produce an enhanced alpha-element to 
iron ratio for the metal-poor halo stars.   The excellent agreement seen in 
Figure~1 between the ICM data and the stellar data rather leads us here to 
propose that the IMF is indeed
invariant, and equal to that of Galactic stars, that the Type II yields 
give enhanced alpha-elements, and that the
ICM {\it did\/} undergo substantial enrichment by Type Ia supernovae. 
Detailed models of the chemical enrichment of the ICM are outwith the scope 
of this {\it Letter\/}, but reduction of a Type~II value of [O/Fe]$\sim 
+0.3$ to solar ratios obviously can be achieved by adding an equal amount of 
iron (see also Ishimaru and Arimoto 1997 and Gibson {\it et al.} 1997), 
consistent with the location of the ICM data in Figure 1.

An alternative way of looking at overall enrichment is afforded by 
the elemental mass-to-light ratio, 
calculated (Loewenstein and Mushotzky 1996) 
as the mass of a given element in the ICM, normalised by the
starlight in elliptical galaxies in the cluster; this approximates the ratio
of mass synthesized and ejected, to mass locked up in low mass stars
(assuming the stars in elliptical galaxies are the source of the
enrichment).  Loewenstein and Mushotzky (1996) found that with standard Type 
~II yields and a Salpeter IMF, the typical iron mass-to-light ratio observed 
(0.02 in solar units; Arnaud {\it et al.} 1992) was 
a factor of two above the pure Type~II enrichment predictions (note that in
this comparison, the absolute value of solar iron used should be irrelevant
as long as the same value was used for both observations and predictions;
see also Ishimaru and Arimoto 1997). This discrepancy  is removed with the
additional Type Ia contribution implied by the element ratios. 

Nevertheless, it should be noted that 
the upper end of the derived range
of blue-luminosity-normalised 
ICM abundances remains higher than the predictions of the yields and IMF 
discussed here. For example, as discussed by Loewenstein and Mushotzky 
(1996), AWM~7 has an estimated iron mass-to-light ratio  
that is three times higher 
than this typical value, and the lower limits on the 
oxygen mass-to-light ratio they derive for the clusters 
Abell~496 and AWM~7 are a factor of two higher than the Salpeter IMF 
predictions.  Thus  some 
parameterizations of the 
elemental abundances in some clusters remain puzzling.  
However, 
all the quantities involved are rather uncertain -- for example, the
contribution of galaxies other than giant ellipticals was ignored, and
there have been suggestions that dwarf galaxies may contribute significantly to
the ICM enrichment (e.g. Trentham 1994)
as could spiral galaxies.  Indeed, the hierarchical-clustering model of
Kauffmann and Charlot (1997), in which ellipticals form by the merging of
smaller galaxies, predicts that around 75\% of the metals in the
intracluster medium of a present-day rich cluster (with circular velocity of
1000km/s) were ejected by disk galaxies with circular velocity less than 
250km/s, not all of which have merged into ellipticals.

\section{Conclusions}

 An invariant IMF is suggested by a wide range of observations, however
 remarkable this may be. Simplicity is appealing, and we here argue
 that the chemical abundances in the Intra Cluster Medium in rich
 clusters of galaxies may also be fit acceptably by the IMF that
 provides an understanding of Galactic stars.

\acknowledgements 

I thank all at the Center for Particle Astrophysics (UC Berkeley) for their
hospitality during the writing of this paper, and the organisers of
the IMF meeting in Cambridge (UK) that motivated it.  I also thank the 
referee, and 
Gerry Gilmore, Joe Silk and Mike Loewenstein 
for comments and discussions.  
This work was
partially supported by  NASA ATP grant NAG5-3928.

\clearpage
\begin{figure}
\plotone{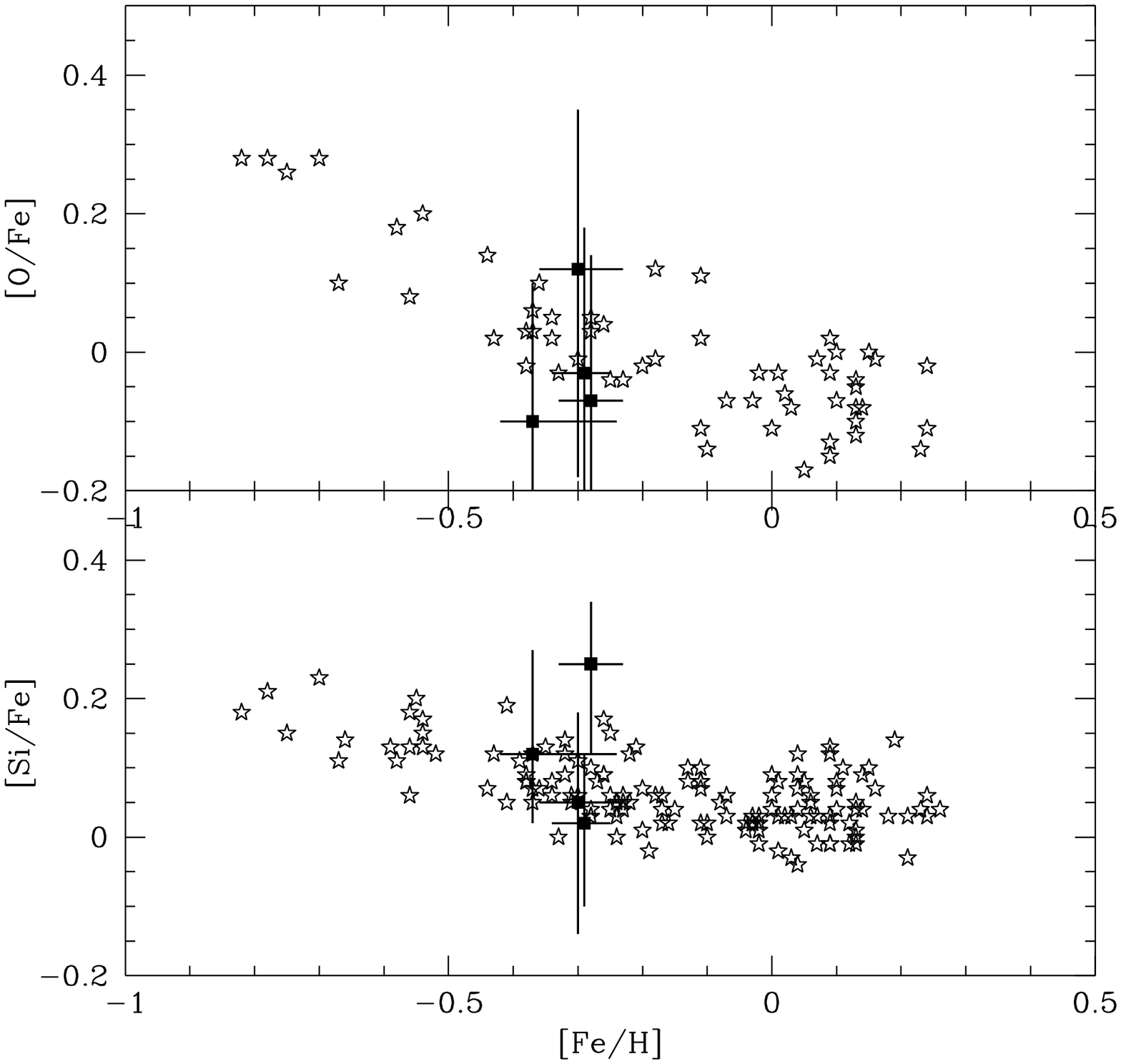}
\caption{Element ratios against iron abundance for 
local disk stars in the Milky Way Galaxy (star
symbols; Edvardsson {\it et al.} 1993, their Tables~11 and 12; 
typical uncertainty is 0.05 -- 0.1~dex in all quantities), 
together with the data for the four X-ray  clusters of Loewenstein and 
Mushotzky (1996), renormalized to the same solar iron abundance (solid
squares, with 90\% confidence limits indicated). Top panel for oxygen, lower for
silicon, these elements chosen to facilitate the comparison.}
\end{figure}

\end{document}